\documentclass[a4paper,11pt]{article}
\usepackage{pos}

\usepackage[utf8]{inputenc}

\usepackage{fancyvrb} % Monospace
\usepackage{graphicx} % Figures
\usepackage[font=small,labelfont=bf]{caption} % Small caption fontsize and bold Figure label
\usepackage{xspace}   % xspace for new command definition

\usepackage{amsmath} % math
\usepackage{slashed}

\usepackage{bbold} % identity symbol \mathbb{1}

\usepackage{environ}

\usepackage{framed}

\usepackage{listings}
\input{ltstyle_module-code_input.sty}
\input{ltstyle_module-colors.sty}
\newcommand{\marty}{\protect\Verb+MARTY+\xspace}
\newcommand{\op}{\ensuremath{\hat{\mathcal{O}}}}

\lstnewenvironment{code}
{
}
{
}

\title{The \marty user interface for the calculation of general Wilson coefficients}
\author*[a]{Gr\'egoire Uhlrich}

\affiliation[a]{University of Geneva,
  24 rue du G\'en\'eral-Dufour, 1211 Geneva 4, Switzerland}

\emailAdd{gregoire.uhlrich@unige.ch}

\abstract{
The calculation of one-loop Wilson coefficients for general Beyond the Standard Model (BSM) scenarios is a technical challenge often addressed by doing long and error prone analytical calculations by hand. Several software programs already provide squared amplitude calculations at the loop-level, but few of them are also able to derive general loop-level Wilson coefficients necessary e.g. for the study of quark decays in flavor physics. \marty, a computer program that automates tree-level and one-loop perturbative calculations for general BSM scenarios can in particular be used to obtain such Wilson coefficients. We present in details the simple user interface allowing to derive common Wilson coefficients in \marty, and the most general use case of \marty to extract the coefficient of any effective operator.
}

\FullConference{%
  Computational Tools for High Energy Physics and Cosmology (CompTools2021)\\
  22-26 November 2021\\
  Institut de Physique des 2 Infinis (IP2I), Lyon, France
}

\begin{document}
\maketitle

\section{Introduction}

Automated calculations beyond the Standard Model have always been a challenge, in particular at the loop level. A lot of software development work has been dedicated to this particular issue in the past decades, with the use of symbolic computation frameworks required for this type of theoretical calculations. Open-source codes implementing their own symbolic computation modules exist such as \Verb+LanHEP+~\cite{lanhep} for the vertex derivation from the Lagrangian, or \Verb+CompHEP+~\cite{comphep} and \Verb+CalcHEP+~\cite{calchep} that automate the calculation of tree-level squared amplitudes in a variety of BSM scenarios. Finally, \Verb+MadGraph_aMC@NLO+~\cite{madgraph, madgraph_nlo} is also open-source and provides tree-level and one-loop calculation facilities for squared amplitudes. If Mathematica~\cite{mathematica}, a closed and commercial computer algebra system, can be used, several other packages exist such as \Verb+FeynRules+~\cite{feynrules, feynrules_nlo}, \Verb+FeynArts/FormCalc+~\cite{feynarts, formcalc} that make use of \Verb+FORM+~\cite{FORM}, \Verb+SARAH+~\cite{sarah} that was initially specialized for supersymmetric (SUSY) models, or packages also dedicated to the Wilson coefficients calculations e.g. \Verb+FormFlavor+~\cite{formflavor} or \Verb+FlavorKit+~\cite{flavorkit}.

\marty~\cite{marty} is a public C++ program, using its own symbolic computation machinery, automating the calculation of amplitudes, squared amplitudes and Wilson coefficients up to the one-loop level for a very large variety of BSM scenarios. A comprehensive documentation is available for \marty, including manuals~\cite{martymanual, cslmanual} and an interactive \Verb+HTML+ documentation~\cite{martydoc}. All the publicly available material related to \marty (code, publications, talks, documentation, etc) can be found on the website~\cite{martysite}.

The automated analytical calculation of one-loop Wilson coefficients for general BSM scenarios, in particular up to dimension-6 operators (with four fermions), is currently not provided by free-to-use packages other than \marty. In this conference paper, we present the main steps to calculate such quantities in any model that can be built with \marty (for more details about model building in \marty see~\cite{martymanual, marty-tools}). The procedure is fully general and allows users to derive for example the loop-level Wilson coefficients relevant for flavor physics, including for (chromo-)magnetic operators and $\Delta F=1,2$ dimension-6 operators for e.g. $b\rightarrow s$ transitions. Section~\ref{sec:def} introduces definitions important to understand what Wilson coefficients are in \marty, and section~\ref{sec:interface} presents the main features necessary to extract these quantities for any BSM model. While~\cite{martyeps} was mainly focusing on Wilson coefficients for flavor physics giving one complete example, here we highlight the general the general procedure and discuss the extension to general Wilson coefficients.

\section{Definitions in \marty}
\label{sec:def}
\subsection{Generalities}
Wilson coefficients are symbolic scalar expressions in front of operator structures in \marty's amplitudes. In Effective Field Theories (EFT), amplitudes are the matrix elements of an effective Hamiltonian
\begin{equation}
\mathcal{H}_{eff}\equiv \sum _iC_i\op_i,
\end{equation}
with $\op _i$ effective operators and $C_i$ their respective Wilson coefficients. The transition amplitude between an initial state $i$ and a final state $f$ is defined as the matrix element of this Hamiltonian:
\begin{equation}
i\mathcal{M}(i\to f)=\langle f|(-i\mathcal{H}_{eff})|i\rangle = -i\sum _i C_i\langle f|\op|i\rangle.
\end{equation}
The operator matrix elements $\langle f|\op|i\rangle$ may not in general be calculated perturbatively and can contain long distance effects. However, the BSM dependence lies in the Wilson coefficients and a perturbative calculation is enough to determine their respective values as explained in e.g.~\cite{buras}. In \marty, a matrix element is simply a particular contraction of external fields. The general case for an amplitude with $N$ external fields $\{\Phi^{\{A_I\}}_I\}_I$ with indices $\{A_I\}$ can be written as
\begin{equation}
\label{eq:alphadef}
	i\mathcal{M} = -i\alpha\sum _i C_i\cdot T_i^{\{A_1\}\cdots\{A_N\}}\cdot \Phi _1^{\{A_1\}}\cdots\Phi _N^{\{A_N\}}, 
\end{equation}
with $T_i^{\{A_1\}\cdots\{A_N\}}$ all different tensors contracting the external fields to each other in the resulting amplitude and $\alpha$ a convention dependent constant. Therefore, by multiplying the amplitude by $i/\alpha$ the Wilson coefficients can be directly identified in front of the different matrix elements.

\marty can decompose amplitudes in independent external field contractions and give the coefficients in front, taking into account a global user-defined factor $\alpha$. The matrix element (a.k.a operator in \marty) is therefore the contraction of fields in the amplitude (including possible tensor couplings), and the Wilson coefficient is the scalar multiplicative factor in front. The particular cases of dimension-5 and dimension-6 operators are discussed in the following.

\paragraph{LO vs. NLO}

As implicitly stated above, the matching used by \marty is trivial and in particular no explicit calculation is performed in the effective theory. This is because \marty provides automated procedures only for the Leading Order (LO), at tree-level or at the one-loop level. In order to obtain Next-to-Leading Order (NLO) Wilson coefficients (e.g. a one-loop calculation for a process that is non-zero at tree-level) one has to perform the same calculation in the effective theory and match the result on the full theory. Although this can be done with \marty, for now no automated procedure allows us to obtain NLO coefficients. Such a procedure could be developed in the future. For further numerical computations from the Wilson coefficients e.g. applying the Renormalization Group Equations (RGE), dedicated open-source codes already exist such as \Verb+SuperIso+~\cite{superiso1,superiso2,superiso3,superiso4} for flavor physics.

\subsection{(Chromo-)Magnetic operators}
\label{sec:d5}
Dimension-5 operators are defined for two fermions $\psi_1$, $\psi_2$ and a vector boson $B$ as
\begin{equation}
\label{eq:mag}
O_\mathrm{mag} \equiv \left(\bar{\psi_1}(T^A)\sigma^{\mu\nu}\Gamma\psi_2\right)F^{(A)}_{\mu\nu},
\end{equation}
with $(T^A)$ the algebra generator when relevant, $F^{(A)}_{\mu\nu}$ the field strength of $B$ and 
\begin{equation}
\Gamma\in \left\lbrace \mathbb{1},\gamma^5,P_L,P_R\right\rbrace.
\end{equation}
To fully define a magnetic operator, a user therefore only has to choose one element picked in a set of 4 elements. The algebra generator $(T^A)$ does not have to be user defined as its presence is determined by the particle types.

\subsection{4-fermions operators}
\label{sec:d6}
Dimension-6 operators with fermions $\psi_1$, $\psi_2$, $\psi_3$ and $\psi_4$ are defined by operators of the type
\begin{equation}
\label{eq:dim6op}
O_{d=6}\equiv T_{ijkl}\left(\bar{\psi_1}^i\Gamma^A\psi_2^j\right)\left(\bar{\psi_3}^k\Gamma^B\psi_4^l\right),
\end{equation}
with Dirac couplings
\begin{equation}
\label{eq:gammaAB}
\begin{split}
\Gamma^A,\Gamma^B\in \big\lbrace\ \quad& \\
&\mathbb{1},\gamma^5,P_L,P_R,\\
&\gamma^\mu, \gamma^\mu\gamma^5, \gamma^\mu P_L,\gamma^\mu P_R, \\
&\sigma^{\mu\nu}, \sigma^{\mu\nu}\gamma^5, \sigma^{\mu\nu}P_L, \sigma^{\mu\nu}P_R\\
\big\rbrace,\quad&
\end{split}
\end{equation}
with $\Gamma^A$ and $\Gamma^B$ contracting to leave no free Minkowski index. The indices $i$, $j$, $k$, and $l$ in equation~\ref{eq:dim6op} are gauge indices, contracted by $T_{ijkl}$ that can be of four main kinds:
\begin{equation}
\label{eq:Tijkl}
\begin{split}
T_{ijkl} &= \delta_{ij}\delta_{kl},\\
T_{ijkl} &= \delta_{il}\delta_{kj},\\
T_{ijkl} &= \delta_{ik}\delta_{jl},\\
T_{ijkl} &= T^A_{ij}T^A_{kl}.
\end{split}
\end{equation}
To fully define a dimension-6 operator, $\Gamma^A$, $\Gamma^B$ and $T_{ijkl}$ must therefore be provided.

\section{The user interface}
\label{sec:interface}

In this section the user interface to obtain the Wilson coefficients of the operators defined above is presented. Four main steps have to be followed in \marty:
\begin{itemize}
	\item Options setup for the amplitude calculation.
	\item Amplitude calculation, including the decomposition on an operator basis.
	\item Definition of the operator of which the coefficient must be extracted.
	\item Extraction of the coefficient.
\end{itemize}

Considering e.g. a process $\psi_1\rightarrow\psi_2 B$ and a \marty model in the \Verb+model+ variable, the two first steps can be performed using:
\begin{framed}
\begin{code}
    FeynOptions options;
    Expr factor = ...; // Convention-dependent factor to be defined if needed
    options.setWilsonOperatorCoefficient(factor);
    vector<Wilson> wilsons = model.computeWilsonCoefficients(
            OneLoop,
            {Incoming("psi1"), Outgoing("psi2"), Outgoing("B")},
            options);
\end{code}
\end{framed}
For the details on how to define the convention-dependent factor we refer to the user manual~\cite{martymanual}. After the calculation, the \Verb+wilsons+ variable contains the decomposed amplitude but work still needs to be done to extract particular coefficients from the result as explained in the the next sections.

\paragraph{The fermion ordering option}
For 4-fermion operators, the order of external fermions in the operator basis must be user-defined. From the initial order given when defining the external particles of the calculation, the final order is defined as a permutation of the initial order. Considering a four fermion process $\psi_1\to\bar{\psi}_2\psi_3\psi_4$, a fermion order $(2, 0, 3, 1)$ corresponds to operators of the type
\begin{equation}
(\bar{\psi_3}\Gamma^A\psi_1)(\bar{\psi_4}\Gamma^B\psi_2),
\end{equation}
where $\Gamma^{A,B}$ are generalized couplings. The indices are defined starting from $0$, a valid permutation is therefore a permutation of $(0, 1, 2, 3)$. The fact that particles are incoming or outgoing is not relevant for this ordering.
Such orderings have to be defined in the options before the amplitude calculation. In the example above the following option must be defined
\begin{framed}
\begin{code}
    options.setFermionOrder({2, 0, 3, 1});
\end{code}
\end{framed}

\subsection{Operator definition}

For common operators, built-in functions exist to create them without having to explicitly construct their explicit analytical expression.\footnote{General operators can also be defined explicitly, see the user manual~\cite{martymanual}.} This is the case for magnetic dimension-5 operators and dimension-6 operators with 4 fermions. 

In order to easily define all possible operators for $d=5$ and $d=6$ discussed in section~\ref{sec:d5} and~\ref{sec:d6} respectively, the different Dirac and color couplings are stored in enumerations. These enumerations are presented in tables~\ref{tab:dim6d} and~\ref{tab:dim6c} respectively.

\begin{table}[h!]
	\centering
	\begin{tabular}{| l | c | c |}
		\hline
		Enumeration element & Name & Expression\\\hline\hline
		\lstinline!DiracCoupling::S! & Scalar & $\mathbb{1}$\\\hline
		\lstinline!DiracCoupling::P! & Pseudo-scalar & $\gamma^5$\\\hline
		\lstinline!DiracCoupling::L! & Left & $P_L$\\\hline
		\lstinline!DiracCoupling::R! & Right & $P_R$\\\hline
		\lstinline!DiracCoupling::V! & Vector & $\gamma^\mu$\\\hline
		\lstinline!DiracCoupling::A! & Axial & $\gamma^\mu\gamma^5$\\\hline
		\lstinline!DiracCoupling::VL! & Vector left & $\gamma^\mu P_L$\\\hline
		\lstinline!DiracCoupling::VR! & Vector right & $\gamma^\mu P_R$\\\hline
		\lstinline!DiracCoupling::T! & Tensor & $\sigma^{\mu\nu}$\\\hline
		\lstinline!DiracCoupling::TA! & Tensor axial & $\sigma^{\mu\nu}\gamma^5$\\\hline
		\lstinline!DiracCoupling::TL! & Tensor left & $\sigma^{\mu\nu}P_L$\\\hline
		\lstinline!DiracCoupling::TR! & Tensor right & $\sigma^{\mu\nu}P_R$\\\hline
	\end{tabular}
	\caption[Dirac couplings for dimension-6 operators]{\label{tab:dim6d}Dirac couplings available to define operator structures in \marty.}
\end{table}
\begin{table}[h!]
	\centering
	\begin{tabular}{| l | c | c |}
		\hline
		Enumeration element & Name & Expression\\\hline\hline
		\lstinline!ColorCoupling::Id! & Identity & $\delta_{ij}\delta_{kl}$\\\hline
		\lstinline!ColorCoupling::Crossed! & Crossed & $\delta_{il}\delta_{kj}$\\\hline
		\lstinline!ColorCoupling::InvCrossed! & Crossed inversed & $\delta_{ik}\delta_{jl}$\\\hline
		\lstinline!ColorCoupling::Generator! & Generator & $T^A_{ij}T^A_{kl}$\\\hline
	\end{tabular}
\caption[Color couplings for dimension-6 operators]{\label{tab:dim6c}Color couplings possible to define dimension-6 operators in \marty. See equation~\ref{eq:dim6op} for the definition of the indices $ijkl$.}
\end{table}

Using the enumeration presented in table~\ref{tab:dim6d}, the \lstinline!chromoMagneticOperator()! method can be used to build the relevant dimension-5 operators defined in equation~\ref{eq:mag} e.g. for $C_7$ ($b\rightarrow s\gamma$ decay) or $(g-2)_\mu$:
\begin{framed}
\begin{code}
    vector<Wilson> O_7 = chromoMagneticOperator(
        model, wilsons, DiracCoupling::R);
       // Fermion current (sigma P_R)
    vector<Wilson> O_gm2 = chromoMagneticOperator( 
        model, wilsons, DiracCoupling::S);
       // Fermion current (sigma)
\end{code}
\end{framed}
A similar principle exists for the dimension-6 operators defined in equation~\ref{eq:dim6op}, this time two Dirac couplings must be given to define the two fermion currents:
\begin{framed}
\begin{code}
    vector<Wilson> O_1 = dimension6Operator(model, wilsons, 
    	DiracCoupling::L, DiracCoupling::R); // (P_L)x(P_R)
    vector<Wilson> O_2 = dimension6Operator(model, wilsons, 
    	DiracCoupling::VL, DiracCoupling::V); // (G^mu P_L)x(G_mu)
\end{code}
\end{framed}

When a gauge tensor coupling of a dimension-6 operator is not trivial, it is possible to specify another one giving the gauge group name (\lstinline!"C"! for color group in the example) and an element of the enumeration presented in table~\ref{tab:dim6c}:
\begin{framed}
\begin{code}
    vector<Wilson> O1_crossed = dimension6Operator(model, wilsons,
    	DiracCoupling::L, DiracCoupling::R,
    	{"C", ColorCoupling::Crossed}
    	); // (P_L)_ij x (P_R)_ji
\end{code}
\end{framed}

\subsection{Wilson coefficient extraction}
Finally, after calculating the amplitude and building the relevant operators as previously discussed, the extraction of the final Wilson coefficients is very simple using the \lstinline!getWilsonCoefficient()! method:
\begin{framed}
\begin{code}
    Expr C7 = getWilsonCoefficient(wilsons, O_7);
    Expr gm2 = getWilsonCoefficient(wilsons, O_gm2);
    Expr C1 = getWilsonCoefficient(wilsons, O_1);
    Expr C2 = getWilsonCoefficient(wilsons, O_2);
    Expr C1p = getWilsonCoefficient(wilsons, O1_crossed);
\end{code}
\end{framed}
The resulting variables are simple \marty symbolic expressions and can therefore directly be used for library generation and numerical evaluation as usual.\footnote{For more details on this procedure see the simple example on the website \url{https://marty.in2p3.fr/gettingStarted.html} or the user manual~\cite{martymanual}.}

\subsection{Generalization of the operator definition}

In \marty it is also possible to extract Wilson coefficients of generic operators. The principle is to create the analytical expression, in \marty, of the operator of which the coefficient must be extracted. Then, \marty automatically searches in the amplitude for the user-defined operator and extracts its coefficient. The creation of custom effective operators is very similar to the creation of general Lagrangian terms and should feel familiar for a user already accustomed to model building procedures in \marty. 

The momenta of a process need in general to be obtained from \marty to define operators. This can be done using:
\begin{framed}
\begin{lstlisting}
    vector<Tensor> p = wilsons.kinematics.getOrderedMomenta();
    // p[0], p[1], p[2] are momenta p1, p2, p3 for a three particles process
\end{lstlisting}
\end{framed}

To obtain the particles, tensors and indices, the procedure is the same as for model building. In order to create gauge indices in the relevant vector spaces, it is necessary to specify the group (or its name) and the irreducible representation (or a particle). For the example of a $b\rightarrow s\gamma$ process this gives:
\begin{framed}
\begin{lstlisting}
    // The fields
    Particle b = model.getParticle("b");
    Particle s = model.getParticle("s");
    Particle A = model.getParticle("A");
    // Additional tensors
    Tensor gamma = dirac4.gamma;
    // Index for the triplet (e.g. quark "b")
    // in the SU(3) color group "C":
    Index i  = model.generateIndex("C", "b"); 
    // Dirac and Minkowski indices
    vector<Index> al = DiracIndices(2);
    Index mu = MinkowskiIndex();
\end{lstlisting}
\end{framed}

Once all objects have been retrieved from \marty, the operator expression can be built explicitly in a symbolic \marty expression. Considering the example of the $\bar{s}(p_2)\slashed{A}(p_3)b(p_1)$ operator, the corresponding \marty expression is:\footnote{For signed indices such as Minkowski indices, it is necessary to specify if the index given to \marty is up or down. By default \Verb+mu+ is down (e.g.\ $A_\mu$) and \Verb!+mu! is up (e.g.\ $A^\mu$).}
\begin{framed}
\begin{lstlisting}
    Expr Op = GetComplexConjugate(s({i, al[0]}, p[1]))*A(mu, p[2])
    		*gamma({+mu, al[0], al[1]})*b({i, al[1]}, p[0]);
\end{lstlisting}
\end{framed}

Finally, the corresponding coefficient can be extracted by \marty giving the operator previously constructed:
\begin{framed}
\begin{lstlisting}
    Expr C = getWilsonCoefficient(wilsons, Op);
\end{lstlisting}
\end{framed}

This procedure is completely general, is valid for all groups and representations, and allows users to extract the Wilson coefficients of all operators that have not been explicitly implemented in the simple user interface for $d=5$ and $d=6$ operators. 

\section{Conclusion}

We presented the user interface in \marty allowing one to extract in a simple way the Wilson coefficients of $d=5$ and $d=6$ operators at the one-loop level. These coefficients are necessary for the calculation of phenomenogically-motivated quantities such as e.g. $(g-2)_\mu$ or the coefficients of quark decays in flavor physics. Furthermore, we showed how to generalize the Wilson coefficient extraction to any effective operator using a procedure similar to the Lagrangian construction in \marty.

We defined analytically the operators in section~\ref{sec:def} by highlighting their specificity and the minimum quantity of information required from a user to uniquely define them. Section~\ref{sec:interface} then presented the interface to build these operators in a \marty program and extract their coefficients for a given process. In a few lines of code, it is possible to extract Wilson coefficients for $d=5$ and $d=6$ operators that can directly be used by the library generation facility of \marty for numerical evaluation. With a reasonable amount of work, the Wilson coefficients of general operators can also be obtained with \marty using the generic operator definition features.

The procedures presented in this proceeding have two very important features:
\begin{itemize}
    \item The code is completely model-independent. Once the procedure is set for the extraction of one or several Wilson coefficients, changing the model in \marty (considering that it has been built) takes only one line and the program will execute in the exact same way. 
    \item Downloading and installing \marty, a free and open-source code, is a sufficient condition to use all the features discussed here.
\end{itemize}
As it has already been showed for beauty quark decays in non-minimal flavor violating MSSM\footnote{Minimal Supersymmetric Standard Model.} scenarios~\cite{nmfv-article}, the user interface for the automated extraction of Wilson coefficients with \marty will greatly facilitate the BSM analyses relying on such theoretical calculations in a large variety of models.

\bibliographystyle{jhep}
\bibliography{biblio}

\providecommand{\href}[2]{#2}\begingroup\raggedright\begin{thebibliography}{10}

\bibitem{lanhep}
A.~Semenov, \emph{{LanHEP: A Package for the automatic generation of Feynman
  rules in field theory. Version 3.0}},
  \href{https://doi.org/10.1016/j.cpc.2008.10.012}{\emph{Comput. Phys. Commun.}
  {\bfseries 180} (2009) 431}
  [\href{https://arxiv.org/abs/0805.0555}{{\ttfamily 0805.0555}}].

\bibitem{comphep}
{\scshape CompHEP} collaboration, \emph{{CompHEP 4.4: Automatic computations
  from Lagrangians to events}},
  \href{https://doi.org/10.1016/j.nima.2004.07.096}{\emph{Nucl. Instrum. Meth.
  A} {\bfseries 534} (2004) 250}
  [\href{https://arxiv.org/abs/hep-ph/0403113}{{\ttfamily hep-ph/0403113}}].

\bibitem{calchep}
A.~Pukhov et~al., \emph{{CompHEP: A Package for evaluation of Feynman diagrams
  and integration over multiparticle phase space}},
  \href{https://arxiv.org/abs/hep-ph/9908288}{{\ttfamily hep-ph/9908288}}.

\bibitem{madgraph}
J.~Alwall et~al., \emph{{The automated computation of tree-level and
  next-to-leading order differential cross sections, and their matching to
  parton shower simulations}},
  \href{https://doi.org/10.1007/jhep07(2014)079}{\emph{Journal of High Energy
  Physics} {\bfseries 2014} (2014) }.

\bibitem{madgraph_nlo}
R.~Frederix et~al., \emph{{The automation of next-to-leading order electroweak
  calculations}}, \href{https://doi.org/10.1007/jhep07(2018)185}{\emph{Journal
  of High Energy Physics} {\bfseries 2018} (2018) }.

\bibitem{mathematica}
{Wolfram Research{,} Inc.}, ``{Mathematica, {V}ersion 12.1}.''
  \url{https://www.wolfram.com/mathematica}.

\bibitem{feynrules}
A.~Alloul et~al., \emph{{FeynRules 2.0 - A complete toolbox for tree-level
  phenomenology}},
  \href{https://doi.org/10.1016/j.cpc.2014.04.012}{\emph{Comput. Phys. Commun.}
  {\bfseries 185} (2014) 2250}
  [\href{https://arxiv.org/abs/1310.1921}{{\ttfamily 1310.1921}}].

\bibitem{feynrules_nlo}
C.~Degrande, \emph{{Automatic evaluation of UV and R2 terms for beyond the
  Standard Model Lagrangians: a proof-of-principle}},
  \href{https://doi.org/10.1016/j.cpc.2015.08.015}{\emph{Comput. Phys. Commun.}
  {\bfseries 197} (2015) 239}
  [\href{https://arxiv.org/abs/1406.3030}{{\ttfamily 1406.3030}}].

\bibitem{feynarts}
T.~Hahn, \emph{{Generating Feynman diagrams and amplitudes with FeynArts 3}},
  \href{https://doi.org/10.1016/s0010-4655(01)00290-9}{\emph{Computer Physics
  Communications} {\bfseries 140} (2001) 418–431}.

\bibitem{formcalc}
T.~Hahn and M.~Perez-Victoria, \emph{{Automatized one loop calculations in
  four-dimensions and D-dimensions}},
  \href{https://doi.org/10.1016/S0010-4655(98)00173-8}{\emph{Comput. Phys.
  Commun.} {\bfseries 118} (1999) 153}
  [\href{https://arxiv.org/abs/hep-ph/9807565}{{\ttfamily hep-ph/9807565}}].

\bibitem{FORM}
B.~Ruijl, T.~Ueda and J.~Vermaseren, \emph{{FORM version 4.2}},
  \href{https://arxiv.org/abs/1707.06453}{{\ttfamily 1707.06453}}.

\bibitem{sarah}
F.~Staub, \emph{{Exploring new models in all detail with SARAH}},
  \href{https://doi.org/10.1155/2015/840780}{\emph{Adv. High Energy Phys.}
  {\bfseries 2015} (2015) 840780}
  [\href{https://arxiv.org/abs/1503.04200}{{\ttfamily 1503.04200}}].

\bibitem{formflavor}
J.A.~Evans and D.~Shih, \emph{{FormFlavor Manual}},
  \href{https://arxiv.org/abs/1606.00003}{{\ttfamily 1606.00003}}.

\bibitem{flavorkit}
W.~Porod, F.~Staub and A.~Vicente, \emph{{A Flavor Kit for BSM models}},
  \href{https://doi.org/10.1140/epjc/s10052-014-2992-2}{\emph{Eur. Phys. J. C}
  {\bfseries 74} (2014) 2992}
  [\href{https://arxiv.org/abs/1405.1434}{{\ttfamily 1405.1434}}].

\bibitem{marty}
G.~Uhlrich, F.~Mahmoudi and A.~Arbey, \emph{{MARTY -- Modern ARtificial
  Theoretical phYsicist: A C++ framework automating theoretical calculations
  Beyond the Standard Model}},
  \href{https://doi.org/10.1016/j.cpc.2021.107928}{\emph{Computer Physics
  Communications} {\bfseries 264} (2021) 107928}.

\bibitem{martymanual}
G.~Uhlrich, ``{MARTY -- User manual}.''
  {\url{https://marty.in2p3.fr/doc/marty-manual.pdf}}, 2021.

\bibitem{cslmanual}
G.~Uhlrich, ``{CSL -- User manual}.''
  {\url{https://marty.in2p3.fr/doc/csl-manual.pdf}}, 2020.

\bibitem{martydoc}
G.~Uhlrich, ``{Documentation of MARTY}.''
  {\url{https://marty.in2p3.fr/doc/marty/html/index.html}}, 2021.

\bibitem{martysite}
G.~Uhlrich, ``{MARTY -- Website}.'' {\url{https://marty.in2p3.fr/}}, 2021.

\bibitem{marty-tools}
G.~Uhlrich, F.~Mahmoudi and A.~Arbey, \emph{{Semi-automated BSM model building
  procedures in MARTY-1.1 through a 2HDM example}},
  \href{https://doi.org/10.22323/1.392.0042}{\emph{PoS} {\bfseries TOOLS2020}
  (2021) 042}.

\bibitem{martyeps}
{G. Uhlrich and F. Mahmoudi and A. Arbey}, \emph{{Automatic extraction of
  one-loop Wilson coefficients in general BSM scenarios using MARTY-1.4}},
  \href{https://arxiv.org/abs/2110.14515}{{\ttfamily 2110.14515}}.

\bibitem{buras}
A.J.~Buras, \emph{{Weak Hamiltonian, CP violation and rare decays}},  in
  \emph{{Les Houches Summer School in Theoretical Physics, Session 68: Probing
  the Standard Model of Particle Interactions}}, Jun, 1998
  [\href{https://arxiv.org/abs/hep-ph/9806471}{{\ttfamily hep-ph/9806471}}].

\bibitem{superiso1}
F.~Mahmoudi, \emph{{SuperIso: A Program for calculating the isospin asymmetry
  of $\ensuremath{{B}\rightarrow {K}^{*}\gamma}$ gamma in the MSSM}},
  \href{https://doi.org/10.1016/j.cpc.2007.12.006}{\emph{Comput. Phys. Commun.}
  {\bfseries 178} (2008) 745}
  [\href{https://arxiv.org/abs/0710.2067}{{\ttfamily 0710.2067}}].

\bibitem{superiso2}
F.~Mahmoudi, \emph{{SuperIso v2.3: A Program for calculating flavor physics
  observables in Supersymmetry}},
  \href{https://doi.org/10.1016/j.cpc.2009.02.017}{\emph{Comput. Phys. Commun.}
  {\bfseries 180} (2009) 1579}
  [\href{https://arxiv.org/abs/0808.3144}{{\ttfamily 0808.3144}}].

\bibitem{superiso3}
F.~Mahmoudi, \emph{{SuperIso v3.0, flavor physics observables calculations:
  Extension to NMSSM}},
  \href{https://doi.org/10.1016/j.cpc.2009.05.001}{\emph{Comput. Phys. Commun.}
  {\bfseries 180} (2009) 1718}.

\bibitem{superiso4}
S.~Neshatpour and F.~Mahmoudi, \emph{{Flavour Physics with SuperIso}},
  \href{https://doi.org/10.22323/1.392.0036}{\emph{PoS} {\bfseries TOOLS2020}
  (2021) 036} [\href{https://arxiv.org/abs/2105.03428}{{\ttfamily
  2105.03428}}].

\bibitem{nmfv-article}
{M. A. Boussejra and F. Mahmoudi and G. Uhlrich}, \emph{{Flavour anomalies in
  supersymmetric scenarios with non-minimal flavour violation}},
  \href{https://arxiv.org/abs/2201.04659}{{\ttfamily 2201.04659}}.

\end{thebibliography}\endgroup

\end{document}